\RequirePackage{fix-cm}
\documentclass[smallextended]{svjour3}       
\smartqed  
\usepackage{graphicx}
\usepackage{comment}
\graphicspath{ {figures/} } 

\usepackage{float}
\usepackage{graphicx}
\graphicspath{ {figures/} }
\usepackage{tabulary}
\usepackage{caption}
\usepackage{siunitx}
\usepackage{array,booktabs,longtable,tabularx}

\usepackage[export]{adjustbox}

\usepackage{footnote}
\usepackage{tablefootnote}

\usepackage{amsmath}
\usepackage{bm}

\usepackage{mathtools,xparse}

\usepackage{amsfonts}
\usepackage{amssymb}
\usepackage{xcolor}

\newcolumntype{P}[1]{>{\centering\arraybackslash}p{#1}}
\newcolumntype{M}[1]{>{\centering\arraybackslash}m{#1}}

\begin{document}

\title{A New Semi-Automated Algorithm for Volumetric Segmentation of the Left Ventricle in Temporal 3D Echocardiography Sequences
}


\author{
	Deepa Krishnaswamy \and
	Abhilash R. Hareendranathan \and 
	Tan Suwatanaviroj \and
	Pierre Boulanger \and
	Harald Becher \and
	Michelle Noga \and
	Kumaradevan Punithakumar \and
}

\institute{D. Krishnaswamy, A. Hareendranathan, P. Boulanger, M. Noga and K. Punithakumar \at 
 		   Department of Radiology and Diagnostic Imaging, University of Alberta, Edmonton, AB, Canada \\
 	       \email{deepa@ualberta.ca}\\
 	       \and
 	       D. Krishnaswamy, A. Hareendranathan, P. Boulanger, M. Noga and K. Punithakumar \at 
 	       Servier Virtual Cardiac Centre, Mazankowski Alberta Heart Institute, Edmonton, AB, Canada
 	       \and
 	       P. Boulanger and K. Punithakumar \at 
 	       Department of Computing Science, University of Alberta, Edmonton, AB, Canada
 	       \and
           T. Suwatanaviroj and H. Becher \at
           ABACUS, Mazankowski Alberta Heart Institute, Edmonton, AB, Canada
}

\date{Received: date / Accepted: date}

\maketitle



\begin{abstract}
\label{sec:abstract}
Purpose: Echocardiography is commonly used as a non-invasive imaging tool in clinical practice for the assessment of cardiac function. However, delineation of the left ventricle is challenging due to the inherent properties of ultrasound imaging, such as the presence of speckle noise and the low signal-to-noise ratio. Methods: We propose a semi-automated segmentation algorithm for the delineation of the left ventricle in temporal 3D echocardiography sequences. The method requires minimal user interaction and relies on a diffeomorphic registration approach. Advantages of the method include no dependence on prior geometrical information, training data, or registration from an atlas. Results: The method was evaluated using three-dimensional ultrasound scan sequences from 18 patients from the Mazankowski Alberta Heart Institute, Edmonton, Canada, and compared to manual delineations provided by an expert cardiologist and four other registration algorithms. The segmentation approach yielded the following results over the cardiac cycle: a mean absolute difference of 1.01 (0.21) mm, a Hausdorff distance of 4.41 (1.43) mm, and a Dice overlap score of 0.93 (0.02). Conclusions: The method performed well compared to the four other registration algorithms.


\end{abstract}

\section{Introduction}
\label{sec:introduction}



Echocardiography is a commonly used medical imaging modality for non-invasive assessment of cardiac function. There are several advantages of using ultrasound imaging such as no ionizing radiation, cost-effectiveness, and high temporal resolution. An advantage of three-dimensional (3D) ultrasound imaging over two-dimensional (2D) imaging is that there is no foreshortening, which is when the ultrasound plane does not pass through the true apex of the heart \cite{Mor-Avi2008}. A second advantage is that for volumetric imaging, 3D ultrasound does not make assumptions about the geometry of the left ventricle (LV) \cite{Mor-Avi2008}. A potential drawback is the lower temporal and spatial resolution and the presence of stitching artifacts \cite{Lang2012,Leung2010_A}.

The ability to diagnose and perform quantitative 3D analysis of the LV is vital for various heart conditions. Common basic clinical measurements include the end-diastolic volume (EDV), the end-systolic volume (ESV), and the ejection fraction (EF) which measures the efficiency of the heart at pumping blood. Clinicians also need to perform advanced analysis over the entire cardiac cycle, as additional measures such as volume and strain can provide complete information about the function of the heart. 

The development of a reliable method to perform delineation of the endocardial borders of the LV to measure the volume and other clinical measures is of vital importance. In current practice, there are several methods often used by clinicians for assessment. One semi-automated method is Simpson's biplane method, which involves the delineation of contours on two long-axis planes  \cite{Lang2005,Lang2015}. Short-axis discs are created automatically between the two contours, and the volume of each disc is found and summed to obtain the final volume of the chamber. A slight modification of this approach, known as the method of discs, requires the user to delineate the contours in the long-axis planes, as well as the short-axis planes, \cite{Lang2005,Lang2015}. One drawback is the dependence on an ellipsoidal geometrical prior for the shape of the LV for both methods. A second disadvantage is the amount of manual interaction, for instance in the method of discs the user is required to delineate the left ventricle on a large number of short-axis slices \cite{Lang2005,Lang2015}.

There have been a number of methods proposed in the literature for volumetric segmentation of the LV in temporal echocardiography sequences. One area of algorithms involves the use of tracking to perform segmentation. For instance, one set of authors \cite{Orderud2006} uses an extended version of the Kalman filter to perform tracking. Several additions have been proposed based on the tracking method \cite{Orderud2006}, where one integrates the use of Doo-Sabin subdivision surfaces \cite{Doo1978} into the extended Kalman filter formulation \cite{Orderud2008}. Others incorporate edge detection into the tracking formulation \cite{Dikici2010,Smistad2014}.

A second set of methods for temporal 3D segmentation involve the use of dynamical appearance models \cite{Huang2012,Huang2014}. A multiscale sparse representation of the appearance is used, where these representations of the appearance are encoded into dictionaries. These dictionaries are then updated as the frames are segmented in the cardiac cycle. Active appearance models (AAM) have also been applied for left ventricular segmentation and tracking, in particular, one set of authors used an AAM for segmentation at the end-diastolic frame, and then performed tracking using the initialization of the previous frame \cite{Stralen2015}. Another method presents a combination of local and global tracking methods based on optical flow and active appearance models \cite{Leung2010_B,Leung2011}.

Active contours have been used in the literature for temporal 3D segmentation, where one set of authors used region-based active contours \cite{Barbosa2012}. An extension was published where tracking is performed by using a 3D optical flow method along with a local block matching method \cite{Barbosa2014}. These methods have been recently extended and improved \cite{Barbosa2017} where both segmentation and tracking have been combined. In one example, an affine motion model from the Lucas-Kanade algorithm is paired with the proposed segmentation framework \cite{Lucas1981,Barbosa2012}. The estimated affine motion can strengthen the temporal information of the segmentations between each of the frames. The optical flow method from the authors \cite{Barbosa2014} has also been extended to use a localized anatomical affine optical flow method \cite{Queiros2017,Pedrosa2017}.

Recently there have been a number of deep learning approaches used for segmentation of the LV in 3D echocardiography data. However, the proposed solutions are for static volumes and not for segmenting the entire cardiac sequence. One approach by a set of authors terms their method VoxelAtlasGAN \cite{Dong2018}, where a conditional generative adversarial network is applied in a voxel-to-voxel basis, and an atlas is used to provide prior knowledge. This approach has been improved in \cite{Dong2020}, where several constraints in the volume consistency and the generative adversarial network have been included in order to improve performance. Another approach termed ACNN, or anatomically constrained neural networks, was developed by authors to also perform 3D segmentation of the LV \cite{Oktay2017}. The method is able to integrate prior geometric knowledge that captures the shape of the LV.

There are several constraints in the methods proposed in the current literature. One issue is the use of an ellipsoidal geometrical prior for the shape of the LV, where the use of a predefined shape could be insufficient for all patients \cite{Orderud2006,Barbosa2012,Barbosa2014}. A second disadvantage is the use of training data, for instance in one method where the initial mesh model is created from a reference mesh \cite{Smistad2014} as well as in the deep learning approaches \cite{Dong2018,Dong2020,Oktay2017}. The algorithms described may also cause artificial changes in the shape of tissues that are not anatomically feasible. 

This study proposes a volumetric segmentation algorithm for the delineation of the LV in temporal echocardiography data. We have focused on segmenting the endocardial contour, though tracking the mid-myocardial contour is more reproducible than the endocardial contour. For MRI analysis and standard datasets, endocardial borders are often traced \cite{Nikitin2006,Bernard2016,Papachristidis2017}, therefore we have focused on the endocardial contours instead of the mid-myocardial ones. The proposed approach has a number of advantages: 1) no dependence on training data; 2) no dependence on geometrical assumptions; and 3) reliance on minimal user interaction. Previous work \cite{Krishnaswamy2018_3D} used only spatial registration to perform 3D segmentation at the end-diastolic and end-systolic phases. This paper extends the formulation by performing diffeomorphic registration in the temporal domain in order to track the cardiac tissue deformation over the entire cardiac cycle, resulting in a 4D segmentation of the chamber. Previous work \cite{Krishnaswamy2018_4D} has described this preliminary formulation of using temporal registration. This paper includes several improvements: 1) use of a larger, varied, patient population 2) a thorough analysis of parameters necessary for the segmentation 3) a detailed and thorough explanation of the methodology.


\section{Materials and Methods}
\label{sec:materials_and_methods}

\begin{figure}
	\begin{minipage}[b]{0.98\linewidth}
		\centering
		\centerline{\includegraphics[width=4.5in, trim=0 2cm 0 0.5cm]{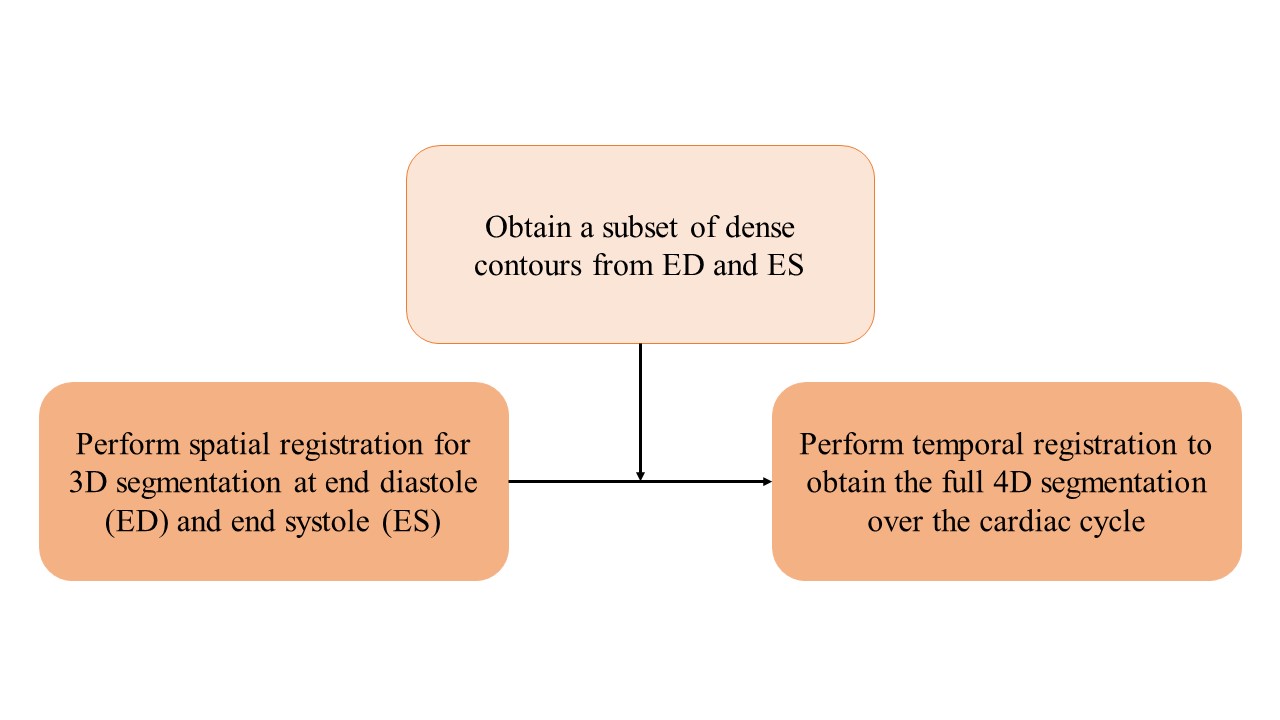}}
	\end{minipage}
	\caption{The overall system diagram for 4D LV segmentation}
	\label{4D_seg_flowchart}
\end{figure}

The flowchart for 4D segmentation of the LV is displayed in Figure~\ref{4D_seg_flowchart}. 3D segmentation is first performed using spatial registration at the end-diastolic (ED) and end-systolic (ES) phases. A subset of these contours at the two phases is then obtained with a set angular spacing. This is followed by a temporal segmentation of the subset of contours across the full cardiac cycle, resulting in a 3D segmentation for each ultrasound frame over the cardiac cycle. 

\subsection{Performing 3D segmentation using spatial registration}
\label{subsec:performing_3D_segmentation}

In order to perform temporal 3D segmentation of the LV over the entire cardiac cycle, the proposed method requires the initial 3D segmentation at the ED and ES frames. The four steps involved in the segmentation process are as follows:  

\begin{enumerate}
	\item Generation of angular slices
	\item Delineation of manual contours
	\item Automatic generation of dense contours
	\item Mesh generation
\end{enumerate}

\subsubsection{Generation of angular slices}
\label{subsubsec:generation_of_angular_slices}

\begin{figure}[htbp]
	\centering
	\includegraphics[width=3.4in]{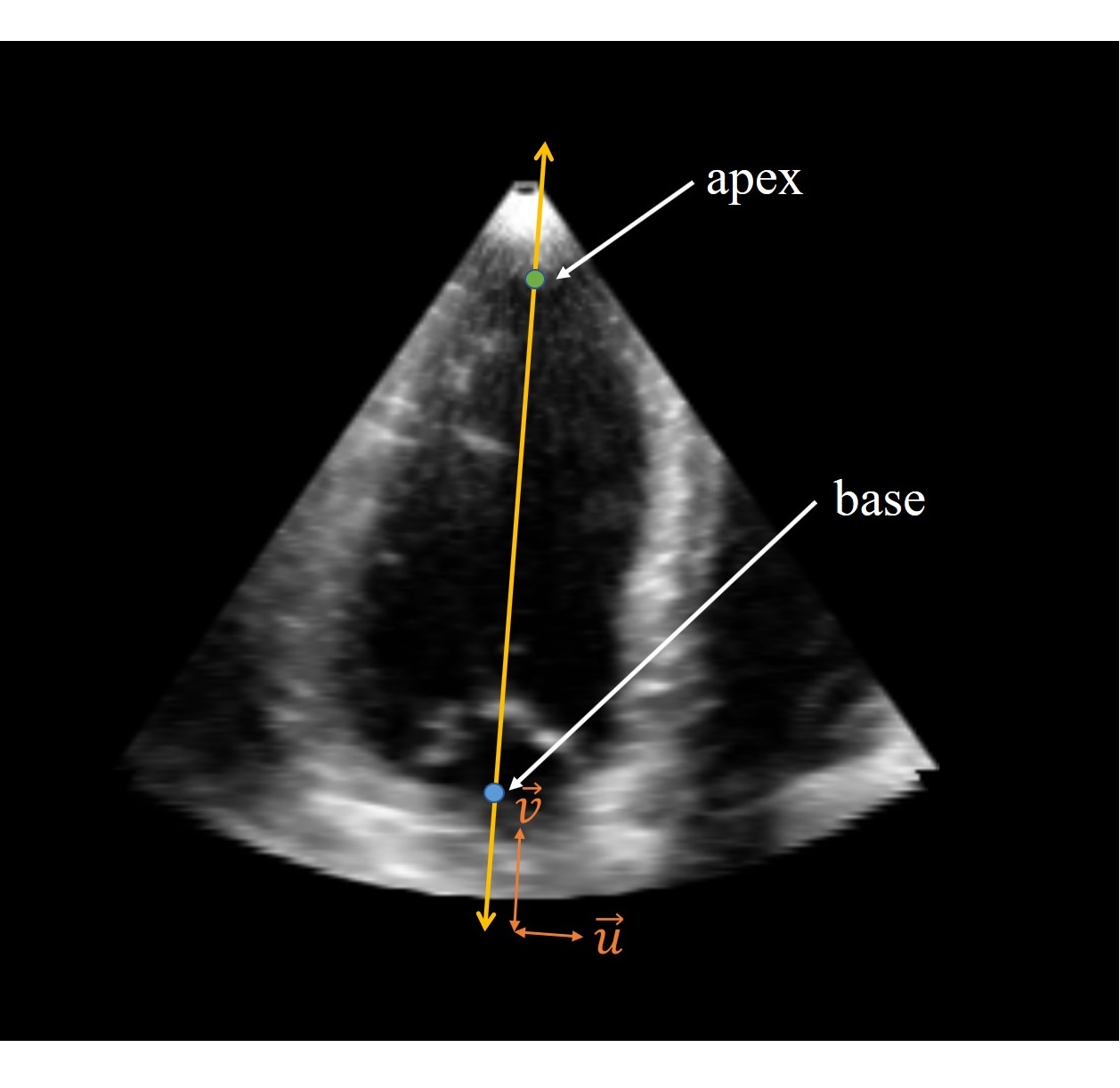}
	\caption{Axis selection for angular slice generation: The yellow line represents the axis, the green point the apex, and the blue point at the base between the mitral valve hinge points.}
	\label{choose_axis_2D_ppt}
\end{figure}

The 3D segmentation approach used involves the generation of angular slices with respect to an axis between the apex and base of the LV. The axis is user-defined and involves choosing a point at the apex, and a secondary point in the center of the mitral valve hinges at the base of the ventricle. Figure~\ref{choose_axis_2D_ppt} displays the process of choosing these two points, in which a 2D plane from a 3D volume is shown. The user-defined axis is shown in yellow, with the point at the apex in green and the point at the center of the base in blue. The same axis is used for both the end-diastolic and end-systolic 3D segmentations.  

\vspace{2cm}

The generation of the 2D angular slices from the user-defined axis is performed using a series of 3D geometric transformations. The first transformation aligns the axis with the $\vec{u} = [1, 0, 0]'$ direction. Define $p_1 \in \mathbb{R}^3$ and $p_2 \in \mathbb{R}^3$ to be the two points that create the user-defined axis between the base and the apex, and let $\vec{v}$ be the unit vector between these two points. Define $\phi$ to be between $\vec{u}$ and $\vec{v}$, and $\vec{r}$ to be $\vec{u} \times \vec{v}$. Rotating $\vec{r}$ by $\phi$ (using the standard rotation matrix from an axis and angle) defines the transformation $T_u$ for aligning the user-defined axis to $\vec{u}$. To automatically slice the 3D data into 2D planes, using the standard rotation matrix about the x-axis, define the second transformation $T_{\theta_{s}}$. The angular slices are created over 180 degrees with an angular spacing of $\theta_d$ degrees. By concatenating the two transformations $T_u$ and $T_{\theta_{s}}$, the final transformation $T_F$ is formed. The origin for performing the angular slicing is defined by $P_{org} \in {\mathbb R}^3$. 

\begin{equation}
T_F = \left[
\begin{array}{cc}
I_3 & -P_{org}\vspace{6pt}\\
0 & 1\
\end{array}	
\right]\times T_{\theta_{s}} \times T_u	\times
\left[
\begin{array}{cc}
I_3 & P_{org}\vspace{6pt}\\
0 & 1\
\end{array}	
\right]	
\label{eq:tf}
\end{equation}
where $I_3$ is the $3\times 3$ identity matrix.

\subsubsection{Delineation of manual contours}
\label{subsubsec:delineation_of_manual_contours}

The 3D segmentation method relies on two contours delineating the endocardial border on two orthogonal slices, which for slices generated over 180 degrees are the $\theta_0$ and $\theta_{90}$ slices. 

\subsubsection{Automatic generation of dense contours}
\label{subsubsec:automatic_generation_of_dense_contours}

\begin{figure}[htbp]
	\centering
	\includegraphics[width=3.4in]{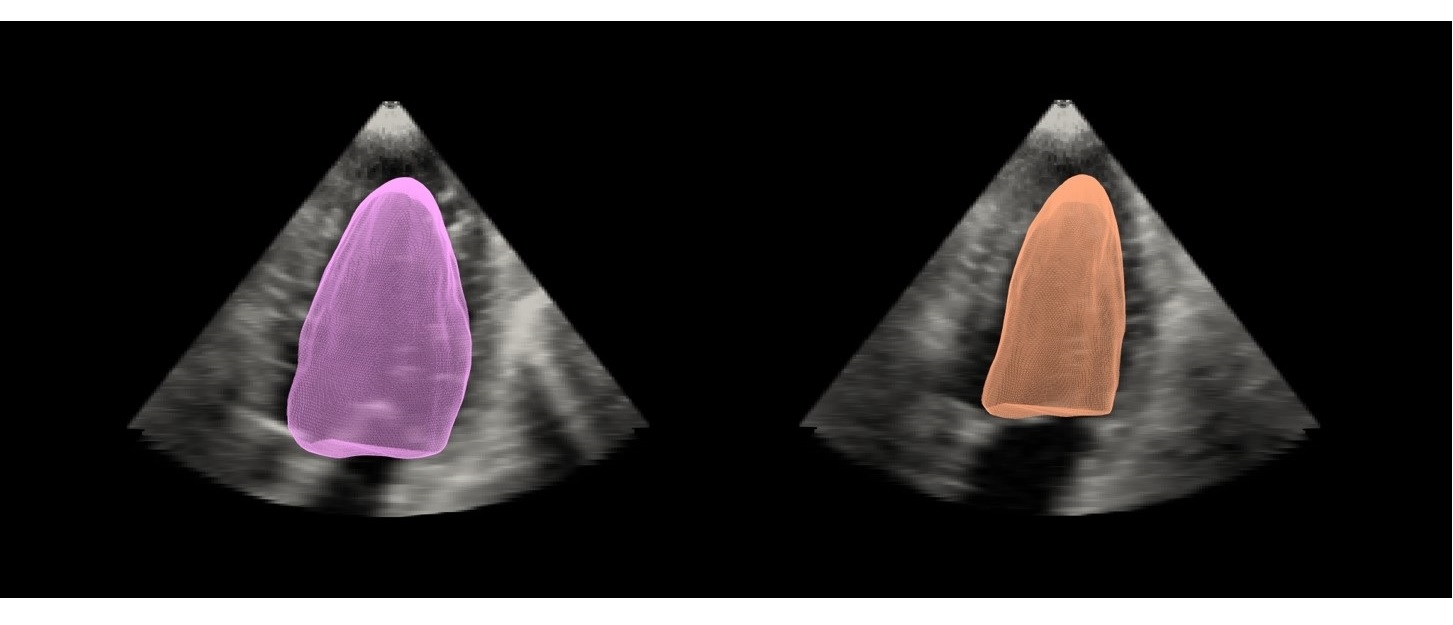}
	\caption{3D contour points showing the delineation of the endocardium in ED and ES using the proposed approach.}
	\label{3d_contour_points_ED_ES}
\end{figure}

Given the two contours on the $\theta_0$ and $\theta_{90}$ slices, a diffeomorphic nonrigid registration algorithm \cite{Punithakumar2017} is used to generate the contours on the other angular slices. The algorithm uses a moving mesh approach, where a point-to-point correspondence is computed between all frames. It is defined using an optimization of a similarity/dissimilarity measure between a pair of frames. The deformation field consists of both radial and rotational components, which are sufficient for capturing the complex motions of the heart. During the optimization process, these radial and rotational components are converted to grid displacements. The method was adapted for use in 2D ultrasound slices in a single 3D volume. Figure~\ref{3d_contour_points_ED_ES} displays the delineation of the endocardium for the ED and ES frames.

\subsubsection{Mesh generation}
\label{subsubsec:mesh_generation}

The automatic generation of the dense contours described in the section above creates a set of 2D contours at each angular slice. To transform the contours to the original space of the ultrasound volume in 3D, the inverse of the geometrical transformations for generating the angular slices was used. A custom program was developed to convert the set of points for each pair of contours to a list of triangle faces necessary for mesh generation exploiting the point-to-point correspondence obtained using the diffeomorphic registration method.

\subsection{Performing 4D segmentation using temporal registration} 
\label{subsec:performing_temporal_3D_segmentation}

A subset of contours for every $\theta_d$ is then extracted from the previously computed ED and ES meshes. In order to obtain the segmentation across the cardiac cycle, the diffeomorphic registration method is applied to the temporal domain for the subset of contours extracted. To increase the temporal consistency of the contours, registration is performed in the forward and reverse directions, and weighting is applied to the contours in order to enforce smoothing. This results in an anatomically plausible mesh at each frame of the cardiac cycle composed of contours at an angular spacing of $\theta_d$ degrees. 

The ground truth meshes at ED and ES were used to automatically choose the appropriate axis and the two $\theta_0$ and $\theta_{90}$ contours. The ground truth meshes were available as they were provided by an expert cardiologist using the TomTec Arena (TomTec Imaging Systems, Unterschleissheim, Germany) clinical software. The apex was chosen to be the point with the minimum z value, and the base point was chosen to be the center of the base of the LV mesh. The two $\theta_0$ and $\theta_{90}$ contours were automatically extracted from the reference mesh based on this axis.

\subsection{Implementation}
\label{subsec:implementation}

The software was implemented using Python 2.7 and the Visualization Toolkit (VTK) package version 5.10.1 \cite{Schroeder2006}. The 3D meshes were created using custom-written software using python and VTK. MeshLab was used to calculate the Hausdorff distance between a pair of meshes \cite{Cignoni2008}.

\section{Results}
\label{sec:results}

The proposed temporal 3D segmentation algorithms were evaluated on ultrasound datasets obtained from $N=18$ adult subjects from the Mazankowski Alberta Heart Institute (Edmonton, Alberta, Canada). The study was approved by the human research ethics committee at the University of Alberta. Subjects were scanned with a Philips iE33 ultrasound machine (Philips Healthcare, Best, the Netherlands) using an X5-1 transducer. A 3D sector angle of $70\times 80$ degrees was optimized in order to achieve a frame rate of greater than 20 volumes per second. The subjects were scanned in order to assess LV function. The number of frames in the cardiac cycle ranged from $17$ to $39$, with an average number of frames equal to $23.83$. The total number of frames for the 18 patients was equal to $421$. The volumes were of dimensions ranging from $160 \times 144 \times 208$ to $256 \times 176 \times 208$ voxels. The voxel resolutions ranged from $0.608 \times 0.787 \times 0.533$ mm to  $0.994 \times 1.339 \times 0.874$ mm. The reference segmentation meshes and volume information were provided by an expert cardiologist from the TomTec Arena software (TomTec Imaging Systems, Unterschleissheim, Germany). 

The results of the 3D spatial segmentation were first analyzed. For the 3D spatial segmentation, distance, overlap and clinical metrics were calculated and compared to the ground truth. A number of experiments were performed to assess the performance of the 3D spatial segmentation algorithm:

\begin{enumerate}
    \item Effect of varying the angular spacing for 3D spatial segmentation 
    \item Effect of the initial axis on the 3D spatial segmentation
    \item Effect of the initial contours on the 3D spatial segmentation
    \item Comparison to a geometrical model
\end{enumerate}

For the 4D temporal segmentation, the proposed method was compared to four other state of the art methods, two variants of the demons algorithm from Insight ToolKit (ITK) \cite{Yoo2002}, optical flow from the Scikit-Image \cite{VanderWalt2014,Zach2007,Wedel2009,Perez2013}, and optical flow from the OpenCV package \cite{Bradski2008,Farneback2003}.

The following distance, overlap, and clinical metrics were calculated to measure the difference between the proposed methods and the ground truth: 

\subsubsection{Mean absolute distance}
\label{subsubsec:mean_absolute_distance}

To calculate the mean absolute distance $d_{m}$ between the proposed approach and ground truth meshes, the average of the distance between each point in the surface in the proposed mesh and the closest point in the reference mesh was calculated \cite{Gottschalk1996}. The result is reported in mm. 

\subsubsection{Hausdorff distance}
\label{subsubsec:hausdorff_distance}

To calculate the Hausdorff distance $d_{H}$, the local maximum distance (Euclidean) is found between the proposed approach $S$ and the reference mesh $R$ using the following equation \cite{Huttenlocher1993}.

\begin{equation}
\text{$d_{H}$}(S,R)  = \max\left\{\adjustlimits\sup_{s\in S} \inf_{r\in R} \mathrm{d}(s,r),  	\adjustlimits\sup_{r\in R} \inf_{s\in S} \mathrm{d}(s,r) \right\}
\end{equation}

The result is reported in mm. 

\subsubsection{Dice score}
\label{subsub:dice_score}

The Dice metric $Dice$ is a measure of overlap between two volumes $V$ and $V_{ref}$, where a value of 1 indicates complete overlap and a value of 0 indicates no overlap \cite{Dice1945}. The $Dice$ score is computed using the following equation. 

\begin{equation}
\text{$Dice$} = \frac{2 (V_s \cap V_r)} {(V_s + V_r)}
\end{equation}

\subsubsection{Ejection fraction}
\label{subsubsec:ejection_fraction}

The ejection fraction indicates the efficiency of the heart at pumping blood. The metric is reported as a percentage, where EDV is the end-diastolic volume and ESV is the end-systolic volume.  

\begin{equation}
\text{EF} = \frac{EDV-ESV}{EDV} \times 100
\end{equation}

\subsection{Quantitative performance evaluation for 3D spatial segmentation}
\label{subsubsec:quantitative_performance_evaluation}

\subsubsection{Quantitative results for end-diastole and end-systole}
\label{subsubsec:quantitative_results_for_end_diastole_and_end_systole}

Table~\ref{distance_metrics_ED_ES} displays the mean absolute difference $d_{m}$ in mm, Hausdorff distance $d_{H}$ in mm, and the Dice score $Dice$ for quantitative evaluation of the accuracy between the proposed segmentation method and the ground truth expert segmentation for the ED and ES frames for all subjects. It can be seen that the ED volume distance measures are in closer agreement to the ground truth meshes compared to the ES measures, indicating the inherent difficulty in capturing the accurate motion of the left ventricle in the ES phase. 

\begin{table}[htbp]
	\caption{Quantitative evaluation results: The mean absolute difference $d_{m}$ in mm, Hausdorff distance $d_{H}$ in mm, and the Dice score $Dice$ for quantitative evaluation of the accuracy between the proposed segmentation method and the ground truth expert segmentation for end-diastole and end-systole for all subjects. The lower the values of $d_{m}$ and $d_{H}$ and the higher the $Dice$, the more accurate the segmentation. The standard deviation values are given in parentheses.}
	\setlength\tabcolsep{8pt} 
	\centering
	\smallskip 
	\begin{tabular}{cccccc}
		\midrule
		\multicolumn{3}{c}{End Diastole} & \multicolumn{3}{c}{End Systole} \\
		\cmidrule(lr){1-3}\cmidrule(lr){4-6}
		$d_{m}$  &  $d_{H}$  &  $Dice$ &  $d_{m}$  &  $d_{H}$  &  $Dice$	  \\
		(mm) & (mm)  &   & (mm)  &  (mm)  &    \\
		\midrule
		0.90 (0.14)   & 4.24 (1.69)   & 0.95 (0.01)  & 0.97 (0.21)   & 4.50 (1.34)   & 0.91 (0.02)    \\  
		\midrule
	\end{tabular}
	\label{distance_metrics_ED_ES}
\end{table}

To assess the quality of the proposed algorithm in terms of clinical measures, the ED volume, the ES volume, and the ejection fraction were calculated for all subjects. To compare to the ground truth values, Bland-Altman analyses were performed for each of the measures \cite{Altman1983}. Table~\ref{clinical_metrics_EDV_ESV_EF} displays the mean difference and standard deviation of the end-diastolic volume, end-systolic volume, and ejection fraction percentage over the subjects. There is a slight bias for underestimation of the ED and ES volumes, and a slightly overestimation of the ejection fraction

\begin{table}[htbp]
	\caption{Clinical metrics: between the proposed method and ground truth segmentations, displaying the mean difference of the end diastolic volume (EDV), end systolic volume (ESV) and the ejection fraction percentage (EF) over the subjects. The standard deviation values are given in parentheses.}
	\setlength\tabcolsep{11pt} 
	\centering
	\smallskip 
	\begin{tabular}{ccc}
		\midrule
		\multicolumn{1}{c}{EDV (mL)} & \multicolumn{1}{c}{ESV (mL)} & \multicolumn{1}{c}{EF (\%)} \\
		\midrule
		4.85 (3.27)   & 2.11 (1.54)  & -0.27 (1.22)   \\  
		\midrule
	\end{tabular}
	\label{clinical_metrics_EDV_ESV_EF}
\end{table}

\begin{figure}[htbp]
    \centering
	\begin{minipage}[b]{0.75\linewidth}
		\centerline{\includegraphics[width=\textwidth,trim={0cm 0cm 0cm 0cm}]{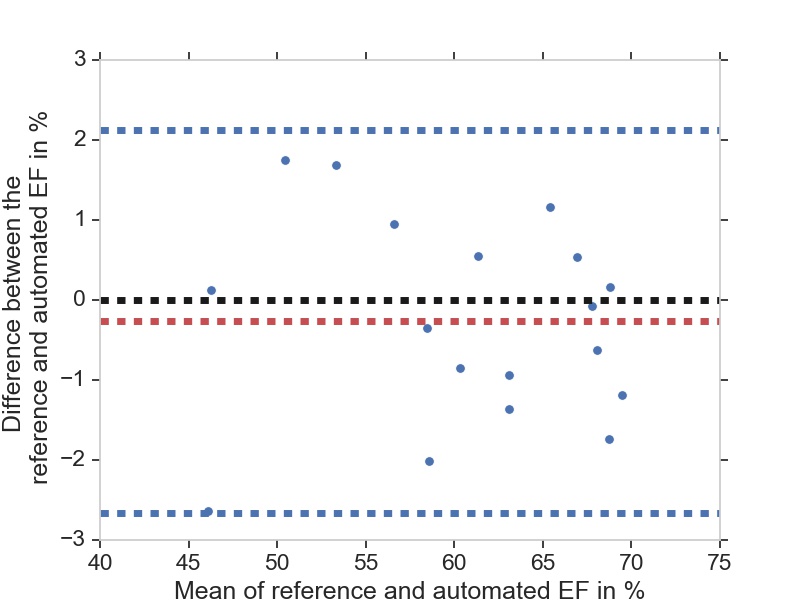}}
	\end{minipage}
	\caption{Bland-Altman plot for the ejection fraction. The black line indicates the zero reference line (perfect agreement), the red line represents the bias, and the two blue lines represent the limits of agreement at two standard deviations away.}
	\label{Bland_Altman_plot}
\end{figure}

The results from the Bland-Altman analysis can also be displayed visually in terms of a plot, where Figure~\ref{Bland_Altman_plot} displays the Bland-Altman plot for the ejection fraction for the subjects. The black dotted line represents the reference line at 0, while the red dotted line represents the bias of the proposed measure subtracted from the reference measure (Table~\ref{clinical_metrics_EDV_ESV_EF}). The dotted blue lines display the limits of agreement, which are $\pm$ two standard deviations away from the bias. It can be seen that only a small bias exists, indicating the ejection fraction was slightly overestimated.

\subsubsection{Effect of varying the angular spacing for 3D spatial segmentation} 
\label{subsubsec:effect_of_varying_the_angular_spacing_for_3D_spatial_segmentation}

\begin{figure}[htbp]
	\begin{minipage}[b]{0.49\linewidth}
		\centering
		\centerline{\includegraphics[width=\textwidth,trim={1cm 0cm 1cm 1cm},clip]{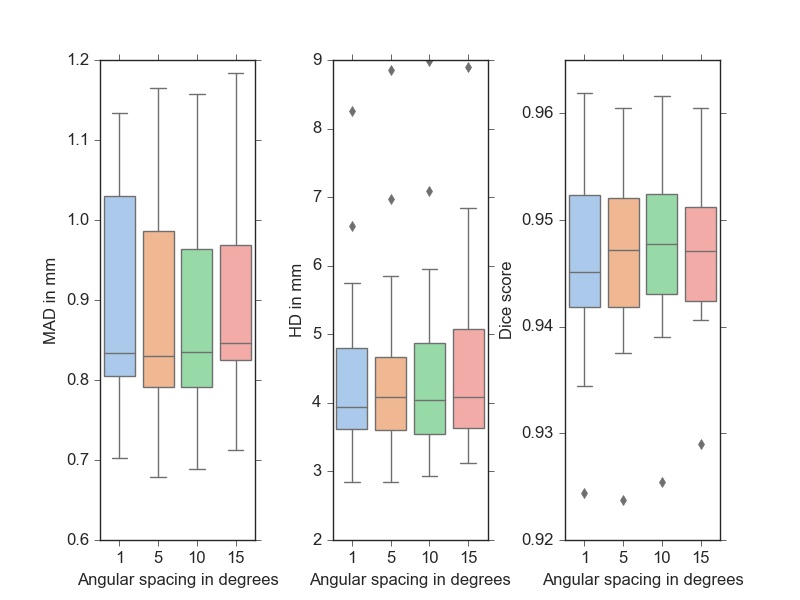}}
		\centerline{\scriptsize End diastole}\medskip 
	\end{minipage}
	\begin{minipage}[b]{0.49\linewidth}
		\centering
		\centerline{\includegraphics[width=\textwidth,trim={1cm 0cm 1cm 1cm},clip]{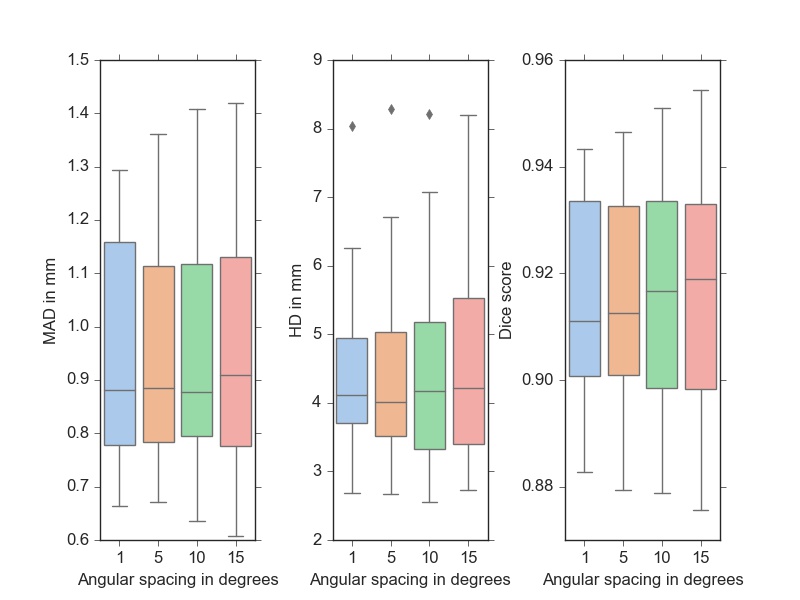}}
		\centerline{\scriptsize End systole}\medskip 
	\end{minipage}
	\caption{Effect of varying the angular spacing for 1, 5, 10 and 15 degrees at end diastole and end systole.}
	\label{angle_spacing}
\end{figure}

One parameter set by the user is the angular spacing $\theta_d$, which determines the total number of contours needed for the individual 3D segmentation for one frame.  With an angular spacing of 1 degree, 180 contours are needed for the 3D segmentation algorithm. By increasing this parameter, one can produce a coarser mesh of the left ventricle and reduce the amount of time needed for the registration process. The angular spacing parameter $\theta_d$  was varied for the ED and ES frames as seen in Figure~\ref{angle_spacing}, using values of 1, 5, 10, and 15 degrees. Qualitatively, it can be observed that for both ED and ES, the evaluation metrics $d_m$, $d_H$ and $Dice$ score do not vary significantly. This is further proved quantitatively by the computation of the Kruskal-Wallis H test, a significance test that determines if the samples are obtained from the same distribution. Table~\ref{angular_spacing_kruskal_wallis} displays the results of the Kruskal-Wallis test for the six parameter combinations ($d_{m}$, $d_{H}$ and $Dice$ scores for the ED and ES volumes). The p values obtained are not below the set significance value of 0.01, indicating that there is not a significant difference between the distance and overlap metrics for any of the angular spacing values. 

\begin{table}[htbp]
	\caption{Kruskal-Wallis H significance tests for the end-diastolic and end- systolic volumes for three distance and overlap metrics: mean absolute distance ($d_{m}$, Hausdorff distance ($d_{H}$) and the Dice score ($Dice$).}
	\setlength\tabcolsep{12pt} 
	\centering
	\smallskip 
    \begin{tabular}{lcccc} 
        \toprule
        & \multicolumn{2}{c}{End-diastole} 
        & \multicolumn{2}{c}{End-systole} \\
        \cmidrule{2-3} \cmidrule{4-5}
        {}   & Test statistic &  p value & Test statistic & p value \\
        \midrule
        $d_{m}$    & 0.305  & 0.959  & 0.014  & 0.999\\ 
        $d_{H}$    & 0.029  & 0.999 & 0.019  & 0.999\\
        $Dice$     & 0.678  & 0.878  & 0.014  & 0.999\\
        \bottomrule
    \end{tabular}
	\label{angular_spacing_kruskal_wallis}
\end{table}

\subsubsection{Effect of the initial axis on the 3D spatial segmentation}
\label{subsubsec:effect_of_the_initial_axis_on_the_3D_spatial_segmentation}

For the 3D spatial segmentation algorithm, the user is required to define the initial axis of the LV. This process may be variable depending on the experience of the clinician, the presence of noise, or the presence of abnormalities. The initial axis was rotated about the +x, +y, +z, -x, -y, -z axes by $\pi/32$ radians for the ED and ES phases. The resulting six meshes were compared to the ground truth in terms of the $d_{m}$, $d_{H}$, $Dice$ and volume in mL. The Kruskal-Wallis test was used in order to determine if there was a significant difference in any of the measures. Table~\ref{initial_axis_kruskal_wallis} displays the results of the Kruskal-Wallis test for the $d_{m}$, $d_{H}$, $Dice$ and volume measures for the original segmentation and the six additional tests. It can be seen that the p values for each measure are all above the set alpha value of 0.01, resulting in non-significant differences for the initial axis test. Therefore, the segmentation of the LV is robust to the selection of the initial axis. 

\begin{table}[htbp]
	\caption{Kruskal-Wallis H significance tests for the initial axis tests for three distance and overlap metrics, mean absolute distance ($d_{m}$, Hausdorff distance ($d_{H}$) and the Dice score ($Dice$), as well as the computation of the $volume$}
	\setlength\tabcolsep{12pt} 
	\centering
	\smallskip 
    \begin{tabular}{lcccc} 
        \toprule
        & \multicolumn{2}{c}{End diastole} 
        & \multicolumn{2}{c}{End systole} \\
        \cmidrule{2-3} \cmidrule{4-5}
        {}   & Test statistic &  p value & Test statistic & p value \\
        \midrule
        $d_{m}$    & 9.155  & 0.165  & 7.501  & 0.277 \\ 
        $d_{H}$    & 1.623  & 0.951  & 0.287  & 0.999 \\
        $Dice$     & 4.988  & 0.545  & 5.262  & 0.511 \\
        $volume$   & 1.206  & 0.977  & 0.186  & 0.999 \\
        \bottomrule
    \end{tabular}
	\label{initial_axis_kruskal_wallis}
\end{table}

\subsubsection{Effect of the initial contours on the 3D spatial segmentation}
\label{subsubsec:effect_of_the_initial_contours_on_the_3D_spatial_segmentation}

The user is also required to define two contours in order to begin the 3D spatial registration process. In order to determine if the algorithm is robust to the selection of the initial contours, the contours were dilated and eroded on a slice by slice basis. The vectors between each point on the contour and the center of the contour were calculated, and dilated or eroded by 1 mm in their respective vector directions. Table~\ref{initial_contour_kruskal_wallis} displays the results of the Kruskal-Wallis H significance tests. It can be seen that for the $d_m$ and $Dice$ metrics, there was a significant difference, but for the $d_H$ and the $volume$ metrics there was not. 

\begin{table}[htbp]
	\caption{Kruskal-Wallis H significance tests for the initial contour tests for three distance and overlap metrics, mean absolute distance ($d_{m}$, Hausdorff distance ($d_{H}$) and the Dice score ($Dice$), as well as the computation of the $volume$}
	\setlength\tabcolsep{12pt} 
	\centering
	\smallskip 
    \begin{tabular}{lcccc} 
        \toprule
        & \multicolumn{2}{c}{End diastole} 
        & \multicolumn{2}{c}{End systole} \\
        \cmidrule{2-3} \cmidrule{4-5}
        {}   & Test statistic &  p value & Test statistic & p value \\
        \midrule
        $d_{m}$    & 27.734  & p$<$0.001  & 16.803  & p$<$0.001 \\ 
        $d_{H}$    & 5.679   & 0.058      & 1.467   & 0.48 \\
        $Dice$     & 27.179  & p$<$0.001  & 13.276  & p$<$0.001 \\
        $volume$   & 3.762   & 0.152      & 3.362   & 0.186 \\
        \bottomrule
    \end{tabular}
	\label{initial_contour_kruskal_wallis}
\end{table}

\subsubsection{Comparison to a geometrical model}
\label{subsubsec:comparison_to_a_geometrical_model}

Some methods that perform segmentation of the left ventricle require the use of a ellipsoid as a geometrical prior. Therefore an experiment was performed to determine if the use of an ellipsoid model produced results that were significantly different from the ground truth. The z axis of the ellipsoid was set to the axis previously used for the spatial segmentation, and the x and y axes were set respectively set to the normals used for the $\theta_0$ and $\theta_{90}$ contours. To determine the size of each of the axes, the z axis was set to the ellipsoid was centered at the point at the base of the mesh. The size of the x and y axes were set to be the distance between the opposing sides of the $\theta_0$ and $\theta_{90}$ contours. This ensured the ellipsoid would fit inside of the two contours. Since this process created a full ellipsoid, larger than the LV itself, it was necessary to cut it at the appropriate plane. The VTK python package was used to cut the ellipsoid automatically, where the z value was chosen to be the highest point of the $\theta_0$ and $\theta_{90}$ contours in the z direction. The cut ellipsoid was then converted to a mesh, where it was then compared to the ground truth using the set of distance and volume measures.  Table~\ref{ellipsoid_model_kruskal_wallis} displays the results of the Kruskal-Wallis H significance tests for the ED and ES meshes. It is apparent that there is a significant difference between the ellipsoid model and the ground truth for all measures except the volume.

\begin{table}[htbp]
	\caption{Kruskal-Wallis H significance tests for the ellipsoid geometrical model test for three distance and overlap metrics, mean absolute distance ($d_{m}$, Hausdorff distance ($d_{H}$) and the Dice score ($Dice$), as well as the computation of the $volume$}
	\setlength\tabcolsep{12pt} 
	\centering
	\smallskip 
    \begin{tabular}{lcccc} 
        \toprule
        & \multicolumn{2}{c}{End diastole} 
        & \multicolumn{2}{c}{End systole} \\
        \cmidrule{2-3} \cmidrule{4-5}
        {}   & Test statistic &  p value & Test statistic & p value \\
        \midrule
        $d_{m}$    & 26.27   & p$<$0.001  & 25.947  & p$<$0.001 \\ 
        $d_{H}$    & 18.515  & p$<$0.001  & 21.926  & p$<$0.001 \\
        $Dice$     & 26.27   & p$<$0.001  & 26.27   & p$<$0.001 \\
        $volume$   & 0.256   & 0.613      & 5.334   & 0.021 \\
        \bottomrule
    \end{tabular}
	\label{ellipsoid_model_kruskal_wallis}
\end{table}

\subsection{Quantitative performance evaluation for 4D temporal segmentation}
\label{subsec:quantitative_performance_evaluation_for_4D_temporal_segmentation}

\subsubsection{Comparison to state-of-the-art methods}

The proposed algorithm was compared to four other registration algorithms, two variants of the demons algorithm from Insight ToolKit (ITK) \cite{Yoo2002}, optical flow from the scikit-image \cite{Zach2007,Wedel2009,Perez2013,VanderWalt2014}, and optical flow from the OpenCV package \cite{Bradski2008,Farneback2003}. The ITK package includes a number of forms of the demons registration algorithm; the classical algorithm and a variant that includes fast symmetric forces used for comparison. For both of the implementations, histogram matching was applied in order to increase the similarity between the two slices being registered. For the classical approach, 50 iterations were used, and 200 iterations were used for the fast symmetric forces method. In order to smooth the displacement field, a Gaussian kernel with a value of 5.0 was used as the standard deviation for both implementations. 

In this version of optical flow from scikit-image \cite{Zach2007,Wedel2009,Perez2013,VanderWalt2014}, a total variation approach using the L1 norm is used, which allows for preservation of discontinuities. An image pyramid approach is also used in a coarse to fine manner in order to improve the optical flow estimation by accounting for large disparities between the images. 
In the OpenCV optical flow implementation, the authors use a method based on polynomial expansion in order to compute the optical flow \cite{Bradski2008,Farneback2003}. The idea is to compute the windows, or neighborhoods around each pixel, by using a quadratic polynomial. By using a polynomial expansion transform, optical flow displacements can be estimated by seeing how this polynomial transforms under translation.  

Table~\ref{metrics_results} displays the results of the 4D distance metrics: the mean absolute difference $d_{m}$ in mm, Hausdorff distance $d_{H}$ in mm, the Dice score $Dice$, and the correlation coefficient of the volumes compared to the ground truth. The results demonstrate high performance of the proposed method compared to the four other registration methods, with a mean $d_{m}$ of 1.01 mm, mean $d_{H}$ of 4.41 mm and a mean Dice score of 0.93.

\begin{table}[htbp]
	\caption{Comparison of the proposed method to four state of the art methods in terms of distance, overlap and clinical metrics: the mean absolute difference $d_{m}$ in mm, Hausdorff distance $d_{H}$ in mm, the Dice score $Dice$, and the correlation coefficient of the volumes in mL compared to the ground truth. The average registration time between a pair of frames is also provided. Values are reported for the subjects over the full cardiac cycle. The standard deviation values are given in parentheses.}
	\setlength\tabcolsep{4pt} 
	\centering
	\smallskip 
	\begin{tabular}{lccccc}
		\midrule
		& \multicolumn{1}{c}{$d_{m}$} & \multicolumn{1}{c}{$d_{H}$} & \multicolumn{1}{c}{$Dice$} & \multicolumn{1}{c}{Corrcoef} & \multicolumn{1}{c}{time}\\
		& \multicolumn{1}{c}{(mm)} & \multicolumn{1}{c}{(mm)} & \multicolumn{1}{c}{} & \multicolumn{1}{c}{} & \multicolumn{1}{c}{(seconds)}\\
		\midrule
        $\quad$ Proposed method             & 1.01 (0.21)  & 4.41 (1.43) & 0.93 (0.02)   & 0.993  & 0.124 \\
		$\quad$ ITK Demons                  & 1.45 (0.49)  & 6.23 (1.48) & 0.89 (0.04)   & 0.969  & 0.452 \\ 
		$\quad$ ITK Demons fsf              & 1.58 (0.53)  & 5.95 (1.70) & 0.89 (0.03)   & 0.972  & 1.921 \\ 
		$\quad$ OpenCV optical flow         & 1.68 (0.52)  & 6.22 (1.48) & 0.87 (0.04)   & 0.975  & 0.021 \\ 
		$\quad$ Scikit image optical flow   & 1.60 (0.52)  & 6.22 (1.48) & 0.88 (0.03)   & 0.972  & 0.403 \\ 
		\midrule 
	\end{tabular}
	\label{metrics_results}
\end{table}

\begin{figure}[htbp]
	\centering
	\begin{minipage}[b]{0.99\linewidth}
		\centerline{\includegraphics[width=\textwidth,trim={0cm 0cm 0cm 0cm},clip]{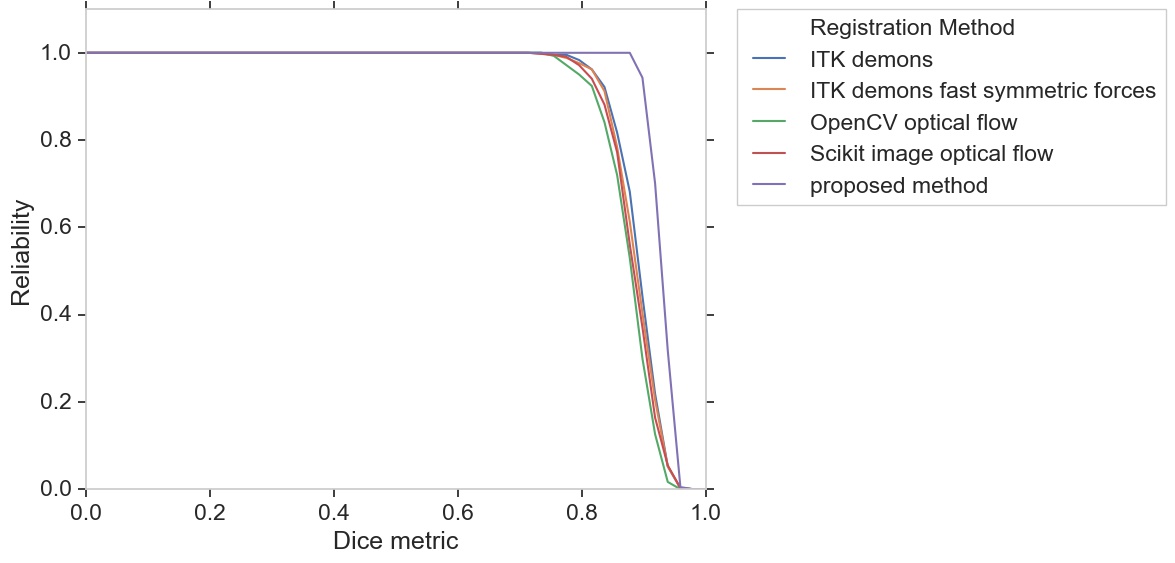}}
	\end{minipage}  
	\caption{Dice reliability plot comparing the four registration methods to the proposed algorithm.}
	\label{Dice_reliability} 
\end{figure}

One approach of comparing the performance of each registration algorithm visually is to use a Dice reliability plot. Threshold values are equally spaced from 0 to 1 (the maximum Dice score), and the percentage of Dice scores that are greater than each threshold value is computed. The Dice reliability plot comparing the proposed method to the four other algorithms is displayed in Figure~\ref{Dice_reliability}. It can be seen that the proposed algorithm includes a larger number of higher Dice scores, indicating better performance compared to the ground truth. 

\begin{figure}[htbp]
	\centering
	\begin{minipage}[b]{0.99\linewidth}
		\centerline{\includegraphics[width=\textwidth,trim={0cm 0cm 0cm 0cm},clip]{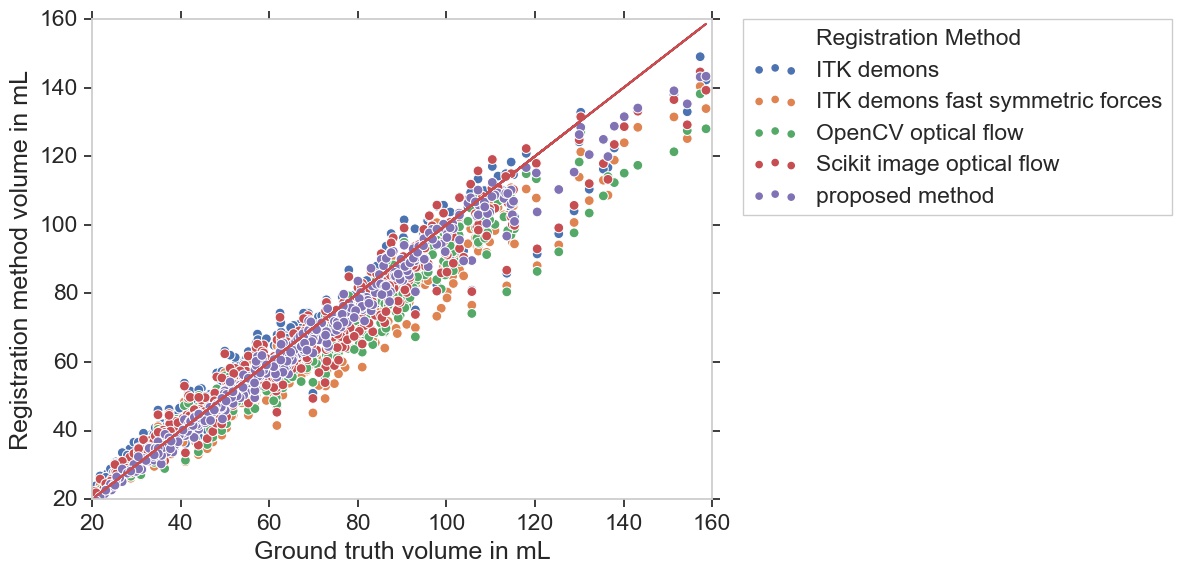}}
	\end{minipage}  
	\caption{Correlation coefficient plot comparing the four registration methods to the proposed algorithm. The red $y=x$ line represents the reference.}
	\label{clinical_corr_plot} 
\end{figure}

Clinical measures are crucial in diagnosing left ventricular function. Obtaining the volume of the chamber in particular is necessary, as it can be used to calculate other metrics such as the ejection fraction. Figure~\ref{clinical_corr_plot} displays the correlation plot of all volumes over the cardiac cycle ($N=421$) for all subjects. The line $y=x$ displays the reference correlation of 1. The correlation plot shows that the proposed method produces volumes that are more clustered around the reference $y=x$ line, indicating better correlation with the ground truth measurements. It can be seen that the other registration methods tend to overestimate the volume. 

\subsection{Visual inspection}
\label{subsec:visual_inspection}

\begin{figure}
	\begin{minipage}[b]{0.49\linewidth}
		\centering
		\centerline{\includegraphics[width=\textwidth]{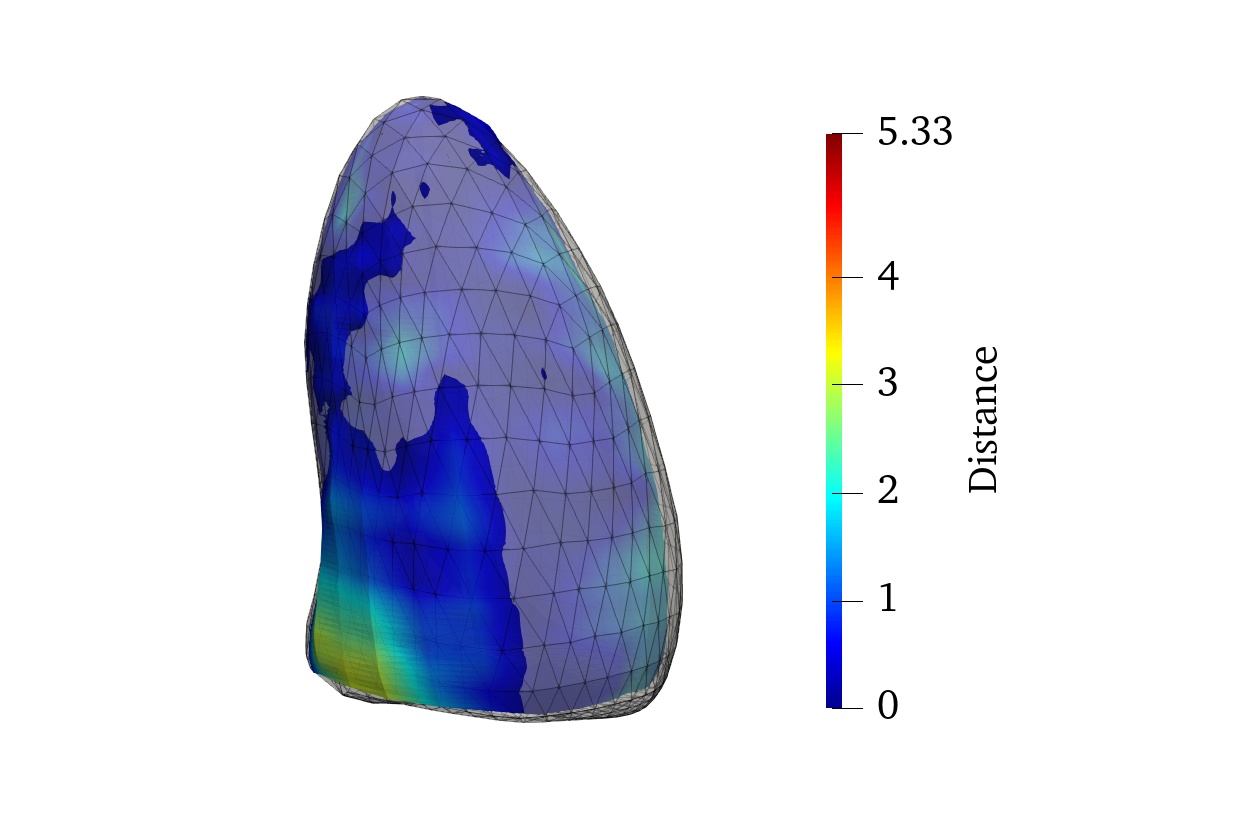}}
		\centerline{\scriptsize (a) End diastolic mesh}\medskip 
	\end{minipage}
	\begin{minipage}[b]{0.49\linewidth}
		\centering
		\centerline{\includegraphics[width=\textwidth]{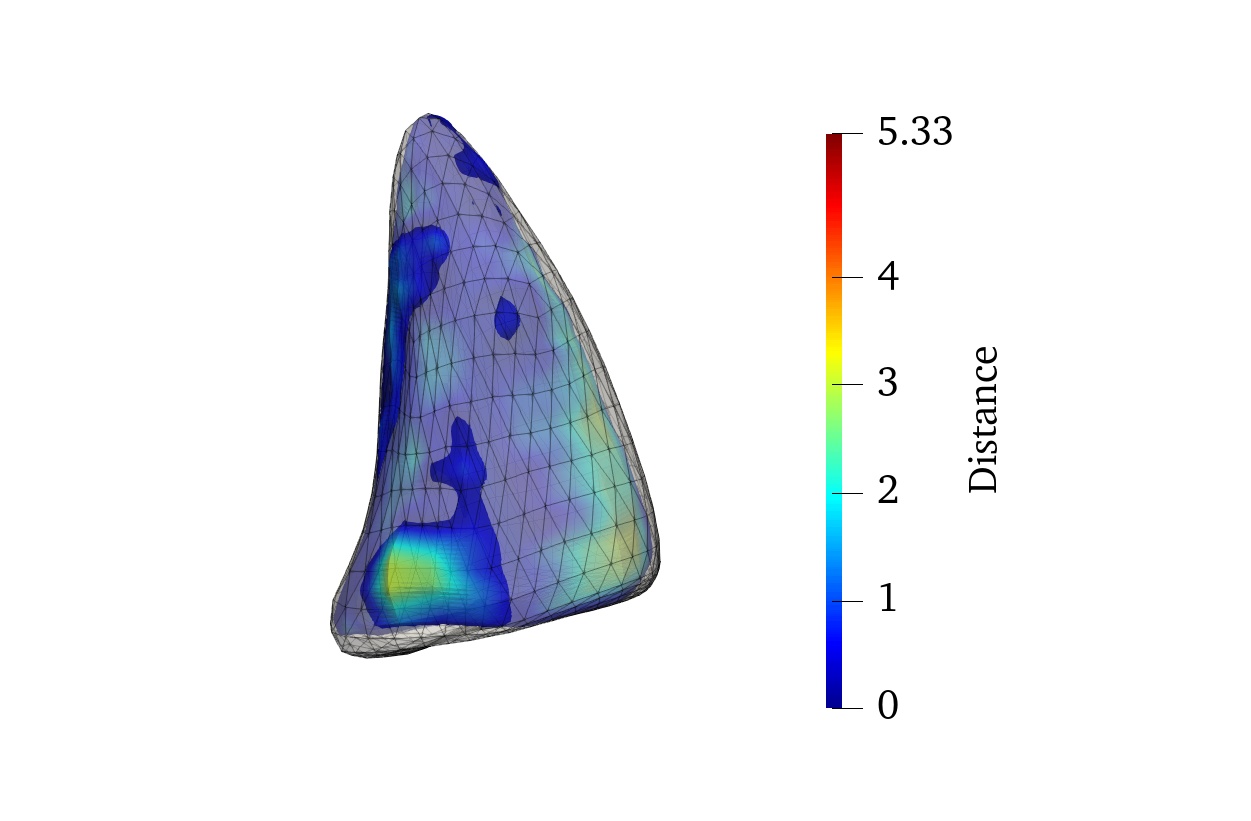}}
		\centerline{\scriptsize (b) End systolic mesh}\medskip 
	\end{minipage}
	\begin{minipage}[b]{0.49\linewidth}
		\centering
		\centerline{\includegraphics[width=\textwidth]{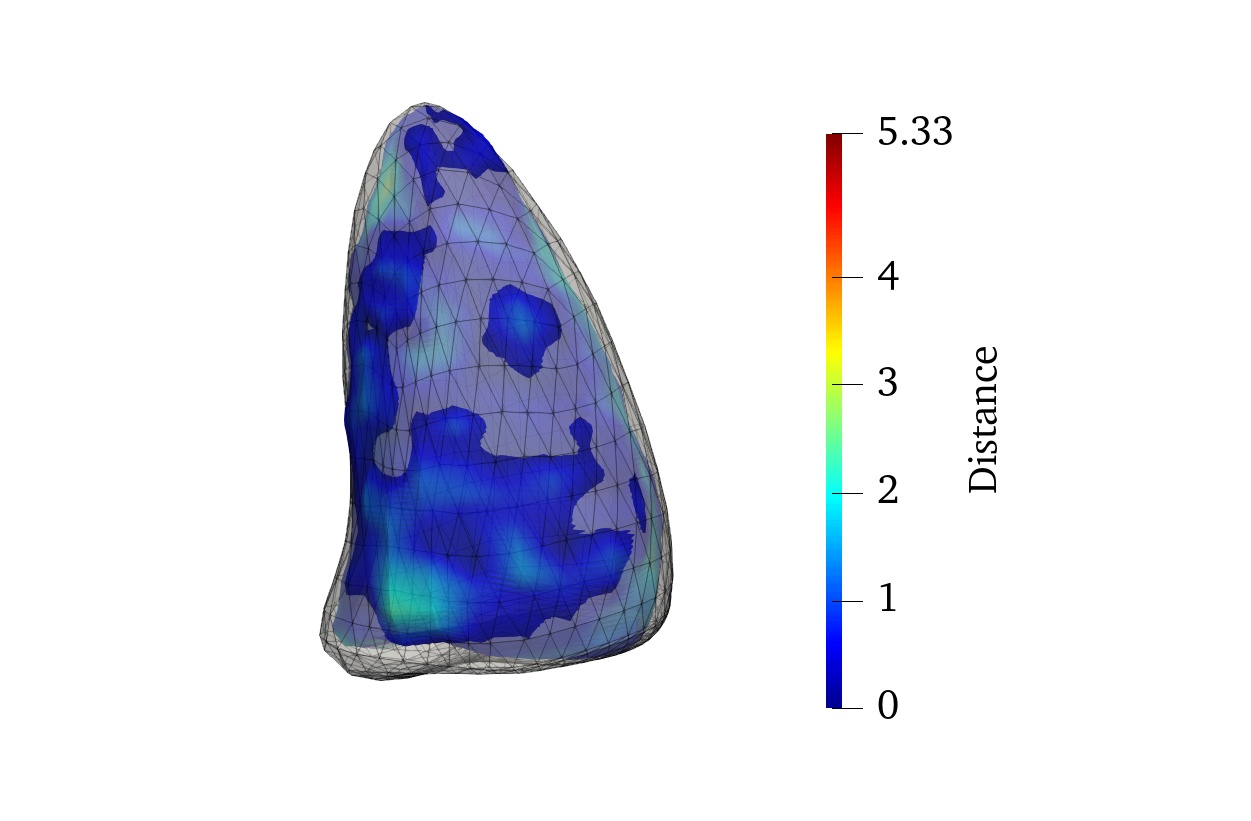}}
		\centerline{\scriptsize (c) Mesh in systolic phase}\medskip 
	\end{minipage}
	\begin{minipage}[b]{0.49\linewidth}
		\centering
		\centerline{\includegraphics[width=\textwidth]{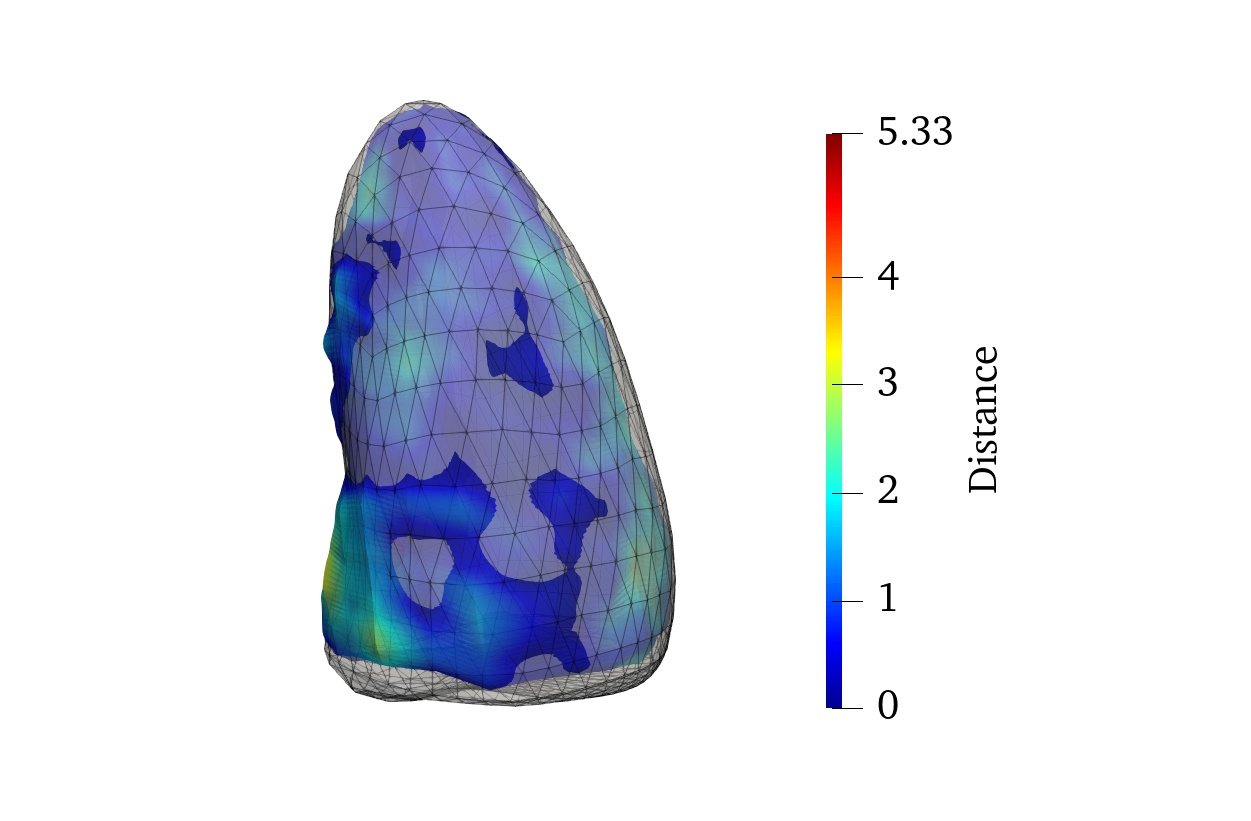}}
		\centerline{\scriptsize (d) Mesh in diastolic phase}\medskip 
	\end{minipage}
	\caption{Comparison of the ground truth mesh in gray and the proposed approach $d_{m}$ mesh in various stages: (a) End diastole (b) End systole (c) Systolic phase (d) Diastolic phase. The color represents the distance values, where blue indicates a small distance from the ground truth, and red indicates a larger difference.}
	\label{visual_inspection_mesh}
\end{figure}

Figure~\ref{visual_inspection_mesh} displays four $d_{m}$ meshes produced by the proposed method overlaid with the ground truth mesh in gray. The four meshes include the ED frame, ES frame, a frame in the systolic phase and a frame in the diastolic phase. The difference is displayed as a heat map, where closer to red indicates a larger difference between the ground truth and the proposed method, and closer to blue, the smaller the distance. The color bars are set to the same range for ease of comparison.  

\section{Discussion} 
\label{sec:discussion}

This study proposes a semi-automated method for segmentation of the left ventricle using temporal sequences of echocardiography data. The proposed algorithm has a number of advantages over current methods available in the literature. One of the main advantages of the algorithm is the inclusion of a diffeomorphic registration approach \cite{Punithakumar2017} for both the spatial and temporal segmentation processes. The diffeomorphic nature ensures that the transform and its inverse exist and are both smooth. This is turn guarantees that grid lines will not cross each other, which is extremely useful in capturing feasible tissue deformation. The use of the radial and rotational components in the diffeomorphic registration method further ensures that true cardiac motion can be represented. Therefore, realistic tissue deformation is modeled and captured by the algorithm, making it especially ideal for cardiac analysis. 

A second advantage of the proposed method is that it does not assume an ellipsoidal shape for the left ventricle. This is crucial as prior geometrical knowledge may not be sufficient for patients with an abnormally shaped left ventricle. Thirdly, training data is not required for the proposed algorithm. Methods that rely on training data (e.g. machine learning and other deep learning approaches) may suffer from collecting data that is varied and also representative of the patient population. Obtaining ultrasound data may also be challenging, as it depends heavily on the vendor and machine settings, and the experience of the clinician. 

There are a number of clinical software packages that are used to perform analysis of the left ventricle, however, several drawbacks exist. One common software used by clinicians is Philips QLab Cardiac Analysis (Philips, Amsterdam, Netherlands) which includes multiple algorithms for analyzing the left ventricle. One algorithm applies Simpson's biplane method, where the user draws two contours in the long-axis slices, and discs are automatically created, and the volume of each disc is calculated and summed. A similar method allows the user to draw the contour for each short-axis slice. These two methods include an assumption about the ellipsoidal shape of the left ventricle. HeartModel is another method provided by Philips that uses a model-based segmentation method. The prior information that is used for the segmentation was obtained from approximately 1,000 echocardiography images. TomTec Arena (TomTec Imaging Systems, Unterschleissheim, Germany) is another clinical software commonly used for left ventricular analysis. In this software, the user first adjusts the orientation of the volume and identifies a predefined set of landmark points. 3D speckle tracking is then performed between the frames in the cardiac cycle in order to perform the left ventricular segmentation. Speckle tracking suffers from lower frame rates, which may cause the incorrect ES frame to be chosen. This in turn may underestimate the calculation of the ejection fraction \cite{Kleijn2012,Yodwut2012}.

The proposed method investigated the performance of the segmentation algorithm at ED and ES, as well as over the cardiac cycle. End diastolic measurements demonstrated high accuracy values with a mean $d_m$ of 0.90 mm and a mean $Dice$ score of 0.95, while end systolic measurements resulted in values of 0.97 mm for $d_{m}$ and a $Dice$ score of 0.91. Clinical measurements also showed reliable values, with a mean ED volume difference of 4.85 mL, ES volume difference of 2.11 mL, and an ejection fraction difference of -0.27\%. 

There are a few limitations to the proposed method for segmentation of the left ventricle. These limitations consist of those concerning the data itself, the semi-automated nature of the algorithm, and the experiments and comparisons performed. In terms of the data itself, one issue is that the quality of the echocardiography images may affect the performance of the diffeomorphic registration method \cite{Punithakumar2017}. The accuracy of the registration method would be improved by including preprocessing of the data, such as speckle noise removal. A limited sample size of $18$ patients was used for the evaluation; in the future this number will be increased. The algorithm was only evaluated on datasets from the same ultrasound machine with frame rates higher than 17 frames per cardiac cycle.

The semi-automated nature of the algorithm poses some limitations. Though it was shown that the proposed approach is robust to the selection of the initial axis, it is not robust in terms of the initial contour selection for the mean absolute distance and Dice score. Therefore, it would be beneficial to develop an automated method of selecting the axis and delineating the contours. One drawback concerning the manual interaction of the proposed method is that it is similar to the amount required by other clinical programs. The current methodology requires the user to choose an axis and delineate two contours on the end-diastolic frame as well as two contours for the end-systolic frame. The TomTec Arena left ventricle analysis software (TomTec Imaging Systems, Unterschleissheim, Germany) requires the user to align the ventricle, and edit the contours at end-diastole and end-systole for the 2, 3 and 4 chamber views. TomTec has a slightly higher manual interaction than the proposed methodology, requiring the annotation of two additional contours. A crucial limitation of the proposed method is the issue of relying on the ground truth to extract the four contours needed from the ED and ES volumes. The comparison to the ground truth is biased as the error is implicitly reduced. In the future, the method will not use the contours extracted from the ground truth, and will instead rely on other machine learning methods.

As for the experiments performed, there exist some limitations. We have fit a single truncated 3D ellipsoid as a geometrical model to the two orthogonal contours for the ED and ES volumes. This is similar to the LV contour initializations from [6, 16, 17] which use a truncated ellipsoid model, or a full ellipsoid that is manually scaled and initialized in the LV cavity. Simpson's biplane method also requires two orthogonal contours, but creates a series of discs from the apex to the base, where a 2D ellipsoid is fit for each of the discs. The volume of each of these discs is found and summed. Unfortunately we did not have access to the actual algorithm, as the Philips QLab Cardiac Analysis (Philips, Amsterdam, Netherlands) software is proprietary; therefore we were unable to compare our method. The software is also unable to save out a mesh of the LV for each frame of the cardiac cycle, which is crucial for the computation of distance metrics. During the spatial registration within a 3D volume, it may be unable to take into consideration large changes in the structure and shape. However, with a small angular spacing value, this rarely occurs. Currently the proposed method was only compared against strictly registration methods, as tracking the points in the LV allows for future regional assessment and motion analysis.

A slightly modified version of the algorithm was also developed for temporal segmentation of the left ventricle. Instead of propagating a subset of the angular contours across the cardiac cycle, the four $\theta_0$ and $\theta_{90}$ contours from the ED and ES frames can be propagated temporally, followed by spatial registration for each frame. This methods resulted in a $d_{m}$ of 1.01 (0.22), $d_{H}$ of 4.59 (1.37) and a $Dice$ score of 0.93 (0.02). Compared to the results from the proposed algorithm, 1.01 (0.21) mm for $d_{m}$, 4.41 (1.43) mm for $d_{H}$ and 0.93 (0.02) for the $Dice$ score, the results from the two segmentation methods are very close. 

Future work includes demonstrating the robustness of the algorithm on a greater number of patients and on a more varied set, such as those with congenital heart disease. The method can also be applied to other chambers of the heart such as the right ventricle, which would be useful because of its complex anatomy. The proposed method can also be applied to segmentation of echocardiography volumes that have been obtained from a multiview fusion approach \cite{Punithakumar2016}.

\section*{Declarations}
\label{sec:declarations}

\subsection*{Funding}
\label{subsec:funding}
The authors would like to thank CIHR/NSERC Collaborative Health Research Projects (CHRP), NSERC Discovery Grant, Heart \& Stroke Foundation of Alberta, NWT, and Nunavut and Servier Canada Inc. for providing the research funding that supported this work. The graphics processor used in this research was donated by the NVIDIA Corporation. 

\subsection*{Conflicts of Interest}
\label{subsec:conflicts_of_interest}
There are no conflicts of interest or competing interests to report. 

\subsection*{Data transparency}
\label{subsec:data_transparency}
The data used for this manuscript is not available. 

\subsection*{Code availability}
\label{subsec:code_availability}
Custom code was used for this manuscript and is not available. 

\subsection{Authors' contributions}
\label{subsec:authors_contributions}

Deepa Krishnaswamy wrote the manuscript, software and algorithms necessary for the project, and performed all of the analysis to obtain the results. Abhilash Hareendranathan provided technical insight and manuscript editing. Tan Suwatanaviroj and Harald Becher provided the ground truth contours for the dataset. Pierre Boulanger provided technical insight and manuscript editing. Michelle Noga provided insight into the methodology, clinical expertise and significant editing of the manuscript. Kumaradevan Punithakumar provided technical expertise, significant editing of the manuscript, including feedback on the figures and tables. 


\section*{Informed Consent}

No animal studies were carried out by the authors for this article. All procedures followed were in accordance with the ethical standards of the responsible committee on human experimentation (institutional and national) and with the Helsinki Declaration of 1975, as revised in 2000 (5). Informed consent was obtained from all patients for being included in the study.



%
%

\end{document}